\documentclass[aps,prl,twocolumn,groupedaddress,showpacs]{revtex4}
\usepackage{amssymb}
\usepackage{amsmath}
\usepackage{cases}
\usepackage{graphicx,color}

\begin{document}


\title{Depth-dependent resistance of granular media to vertical penetration}


\author{T. A. Brzinski III, P. Mayor, and D. J. Durian}
\affiliation{
     Department of Physics \& Astronomy, University of
     Pennsylvania, Philadelphia, PA 19104, USA
}


\date{\today}

\begin{abstract}
We measure the quasi-static friction force acting on intruders moving downwards into a granular medium.  By utilizing different intruder geometries, we demonstrate that the force acts locally normal to the intruder surface. By altering the hydrostatic loading of grain contacts by a sub-fluidizing airflow through the bed, we demonstrate that the relevant frictional contacts are loaded by gravity rather than by the motion of the intruder itself.  Lastly, by measuring the final penetration depth versus airspeed and using an earlier result for inertial drag, we demonstrate that the same quasi-static friction force acts during impact.  Altogether this force is set by a friction coefficient, hydrostatic pressure, projectile size and shape, and a dimensionless proportionality constant.  The latter is the same in nearly all experiments, and is surprisingly greater than one.
\end{abstract}

\pacs{45.70.-n, 47.57.Gc, 83.80.Fg, 81.70.Bt}


\maketitle




The flow of granular systems defies rheological description, in part because shear tends to localize and conventional instruments cannot measure stress and strain \cite{jaeger_granular_1996}.  An alternative approach is to measure the force on a moving intruder~\cite{Schiffer_1999, albert_stick-slip_2001, uehara_low-speed_2003, KoehlerEL05, BehringerPRE05, brzinski_iii_characterization_2009, Gravish_PRL2010, OkumuraEPL10, Schiffer_PRE2011, Ruiz_Review}. For slow horizontal motion at a fixed depth, the force is rate-independent and proportional to both the projected area of the intruder and its depth; this is due to friction acting at gravity-loaded contacts \cite{albert_stick-slip_2001, brzinski_iii_characterization_2009}. For a sphere dropped vertically onto a granular medium, the depth-averaged stopping force, deduced by $\langle F\rangle =mgH/d$ where $H$ is the total drop distance and $d$ is the penetration depth, can be much greater~\cite{uehara_low-speed_2003}.  Most impact behavior can be reconciled by an equation of motion of the form~\cite{tsimring_modeling_2005, katsuragi_unified_2007}
\begin{equation}
	ma = -mg + F(z) + bv^2.
	\label{eq:TsimForce}
\end{equation}
Here $F(z)$ is a rate-independent friction that grows with depth $z$.  And $bv^2$ is an inertial drag proportional to the square of projectile speed $v$, due to momentum transfer just as for a fluid a high Reynolds number.  Ref.~\cite{brzinski_iii_characterization_2009} showed that the stopping force experienced by a horizontally rotating rod is of the same form, but smaller in magnitude, and that $F(z)$ grows with depth due to gravitational loading of frictional contacts. Refs.~\cite{Goldman_PRE2010, Royer_Air} explored the role of air and grain packing fraction. Ref.~\cite{KN_impact} demonstrated that $F(z)$ increases monotonically with the pre-stressing of the packing normal to the direction of gravity. Ref.~\cite{Clark_2012} suggested the existence of an additional constant force term $F_{0}$, finding that both $F_{0}$ and $b$ increase with projectile size. Ref.~\cite{pacheco-vazquez_infinite_2011} demonstrated that $F(z)$ saturates with Janssen-like $z$-dependence for deep impacts.  In the absence of wall effects \cite{pacheco-vazquez_infinite_2011, nelson_projectile_2008, Seguin2008}, the typical assumed form is $F(z)=kz$, dating back to Ref.~\cite{Lohse2004}.

Here we address two outstanding issues with regards to the $F(z)$ friction term in Eq.~(\ref{eq:TsimForce}). First, while the total stopping force points up, the extent to which friction acts locally normal versus tangential to surface area elements of the projectile is not known. Second, while $F(z)$ has been shown to grow with \emph{gravitational} loading of the bed \cite{goldman_scaling_2008, Altshuler_2013}, it is unknown how the \emph{motion} loading of contacts by the intruder affects the total stopping force.  In particular, force chains extend from the projectile deep into the medium, and are intermittently loaded and broken during impact~\cite{karen_2004, pica_ciamarra_dynamics_2004, Clark_2012, Clark_2013}; this could contribute to friction.  Our approach is to directly measure $F(z)$ under two sets of conditions.  In one we vary the shape of the projectile in order to alter the fraction of the projectile surface that moves parallel vs perpendicular to the medium.  In the other, we impose a sub-fluidizing up-flow of air to systematically counteract the gravitational loading of the grains without affecting their motion loading.  One might expect friction between grains and projectile to act tangential to their surface of contact, and to be stronger due to the additional motion loading.  In striking contrast, we find the local friction force to be
\begin{equation}
	{\rm d}{\bf F} = \alpha \mu (\rho g z) {\rm d}{\bf A},
	\label{eq:dFqs}
\end{equation}
where $\mu$ is an internal friction coefficient equal to the tangent of the repose angle, $\rho gz$ is the \emph{gravitational}{ loading pressure, ${\rm d}{\bf A}$ is an infinitesimal area element pointing \emph{normal} to the projectile surface, and $\alpha=35 \pm 5$ is a number we measure to be the same in nearly all experiments.  The total drag force is found by integrating over ${\rm d}{\bf A}$.  Such behavior is relevant for locomotion in and on grains \cite{Goldman_Science2013}, as well as for meteorite strikes \cite{deBruyn_PRL2003, Lohse2004, Ruiz_Review}.

Our granular medium consists of cohesionless glass spheres of diameter range $300\pm50$~$\mu$m; bulk density $\rho = 1.48$~{g/cm$^{3}$}; and draining angle of repose $22^{\circ}$, giving $\mu=0.40$. The grains fill a 19~cm diameter acrylic cylinder to a depth of 20~cm; this is large enough to avoid wall effects \cite{nelson_projectile_2008, Seguin2008}. Underneath is an apparatus for applying a uniform up-flow of air through the granular packing as in Refs.~\cite{katsuragi_unified_2007, brzinski_iii_characterization_2009}.  Before each experiment the grains are first fluidized and the up-flow is then gradually stopped, giving a level surface and a packing fraction of $\varphi=0.59$. 

In the first set of experiments, a projectile is hung from a force gauge suspended beneath a pulley. The projectiles are \emph{cylindrical}, with radii $R$ between 0.476 and 3.85~cm; \emph{conical}, with apex half-angles $\phi$ of 15, 30, 45, and 60~degrees; and \emph{spherical}, with radii $R$ of 1.27~cm, 2.54~cm, and 5.05~cm. The pulley is used to slowly increase the fraction of the projectile's weight that is supported by the granular media. The depth, $z$, of the bottom of the projectile is deduced with a height gauge, and a simultaneous reading of the force gauge is recorded. Initially we observe a single-valued force for each depth, however below a short initial penetration depth we observe stick-slip behavior. In this case, the penetration depth remains constant while the load increases.  Beyond about 20\% \cite{supp}, the granular material fails, and the projectile falls a short distance.  Because $F(z)$ is the rate-independent drag acting on a \emph{moving} intruder, we consider only the forces at which the material fails.

Results for $F(z)$ are plotted against $z$ for fifteen intruders of various sizes and shapes. We normalize according to expectation given by integrating Eq.~(\ref{eq:dFqs}) with ${\rm d}{\bf A}$ pointing \emph{normal} to the surface area elements of the intruder \cite{supp}:
\begin{subnumcases}{\label{eq:Shape} \frac{F(z)}{\alpha\mu\rho g}=}
	\label{eq:Cyl}
		\pi R^{2}z & cylinder \\
	\label{eq:Con}
		({\pi}/{3})\tan^{2}{\phi}~z^{3} & cone \\
	\label{eq:Sph1}
		\pi(R-{z}/{3})z^2 & sphere, $z\leq R$\\
	\label{eq:Sph2}
		\pi(z-{R}/{3})R^2 & sphere, $z\geq R$
\end{subnumcases}	
If ${\rm d}{\bf A}$ points tangential to the area elements, the scaling is quite different: $Rz^2$ for cylinders, $\tan\phi z^3$ for cones, and $(z-0.58R)R^2$ for spheres at $z>R$ \cite{supp}.  Thus we plot (a) $F(z)/R^2$ vs $z$ for cylinders, (b) $F(z)/\tan^2\phi$ vs $z$ for cones, and (c) $F(z)/R^3$ vs $z/R$ for spheres.  Indeed, in accord with Eqs.~(\ref{eq:Shape}) this causes excellent collapse.

\begin{figure}
\includegraphics[width=2.50in]{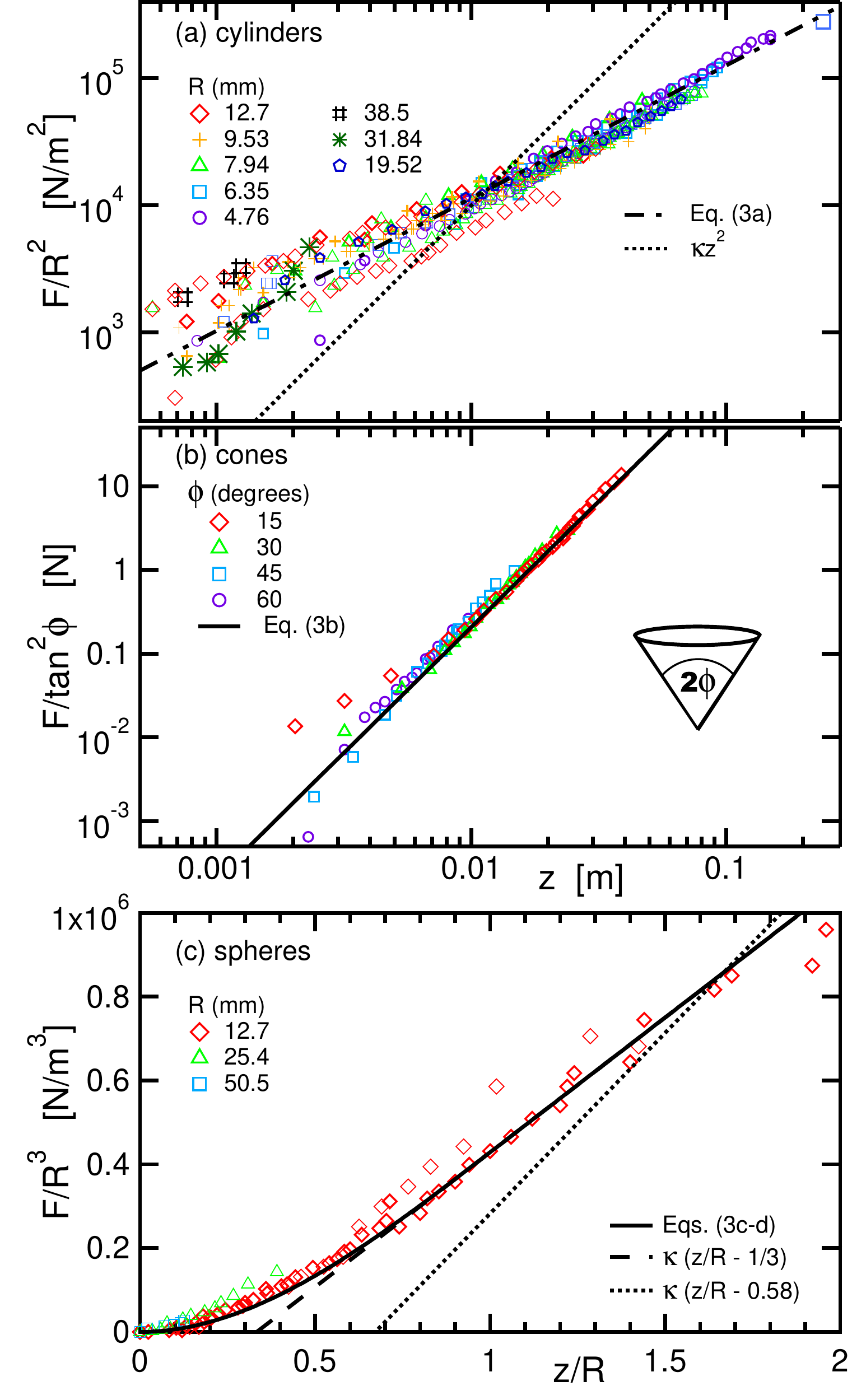}
\caption{(Color online) Friction force versus depth for (a)~cylindrical, (b) conical, and (c)~spherical intruders of different geometry; $R$ is radius and $\phi$ is apex half-angle.  The scaling of the y-axes, and the curves through the data, are the expectations of Eqs.~(3) for ${\rm d}{\bf F}$ normal to the intruder surface area elements.  The dashed (dotted) line in (c) represents expectation for $z>R$ for ${\rm d}{\bf F}$ normal (tangential) to the intruder.}
\label{fig:figure1}
\end{figure}

As a further test, beyond collapse, we compare force data with the expected forms.  In {Figs.~\ref{fig:figure1}(a,b)} for cylinders and cones, the best fit power-laws have respective exponents of $1.06\pm0.06$ and $2.95\pm 0.36$, consistent with a normal direction for ${\rm d}{\bf A}$.  And in Fig.~\ref{fig:figure1}(c) for spheres, the data at $z>R$ fall on a line proportional to $z/R - 1/3$, consistent with Eq.~(\ref{eq:Sph2}) as opposed to the surface-tangent expectation of $z/R - 0.58$ \cite{supp}. For $z<R$ the data are also fit well by Eq.~(\ref{eq:Sph1}). Altogether, the agreement between data and Eq.~(\ref{eq:Shape}) demonstrates that the quasi-static friction force acts primarily \emph{normal} to the area elements of the intruder.  In principle it ought to act tangential, too, but this effect is much smaller.

Next, in order to distinguish the respective roles of gravity-loaded and motion-loaded contacts, we subject the system to a sub-fluidizing flow of air.  Since the airspeed $U$ is proportional to the gradient in air pressure across the sample, this modifies the effective hydrostatic pressure gradient as $\rho g \rightarrow \rho g(1-U/U_c)$ where $U_c$ is the fluidization airspeed where the upflow exactly balances gravity.  Thus, without actually changing gravitational acceleration \cite{goldman_scaling_2008, Schiffer_PRE2011, Altshuler_2013}, the effective gravity loading is reduced by a sub-fluidizing upflow, $0<U<U_c$; and similarly it is enhanced by a down-flow, $U<0$.  The motion-loading of contacts, and the inertial drag force, ought not be affected by airflow.  Note, too, that the very presence of air does not affect the impact behavior since the grains are sufficiently large \cite{katsuragi_unified_2007}. Furthermore, even for smaller grains, air effects vanish near $\varphi=0.58$~\cite{Royer_Air}, which is just below the conditions here of $\varphi=0.59$.

For sample preparation, we begin as usual but then gradually tune the airspeed to a value in the range $U<U_c$.  Downflow, $U<0$, is achieved by connecting the apparatus to a shop-vac.  The rate-independent frictional drag, $F(z)$, is then measured as above.  The results are scaled by geometrical factors, as well as by $(1-U/U_c)$, and plotted in Fig.~\ref{fig:figure2}. For comparison, all data from Fig.~\ref{fig:figure1} are included as grey squares. Again this normalization causes good collapse, and all the data demonstrate the same scaling as the data for $U=0$ presented in Fig.~\ref{fig:figure1}: $F(z)/(1-U/U_c)$ is proportional to $R^2z$ for cylinders, to $\tan^2\phi z^3$ for cones, and is consistent with Eqs.~(\ref{eq:Sph1}-d) for spheres.

\begin{figure}
\includegraphics[width=2.50in]{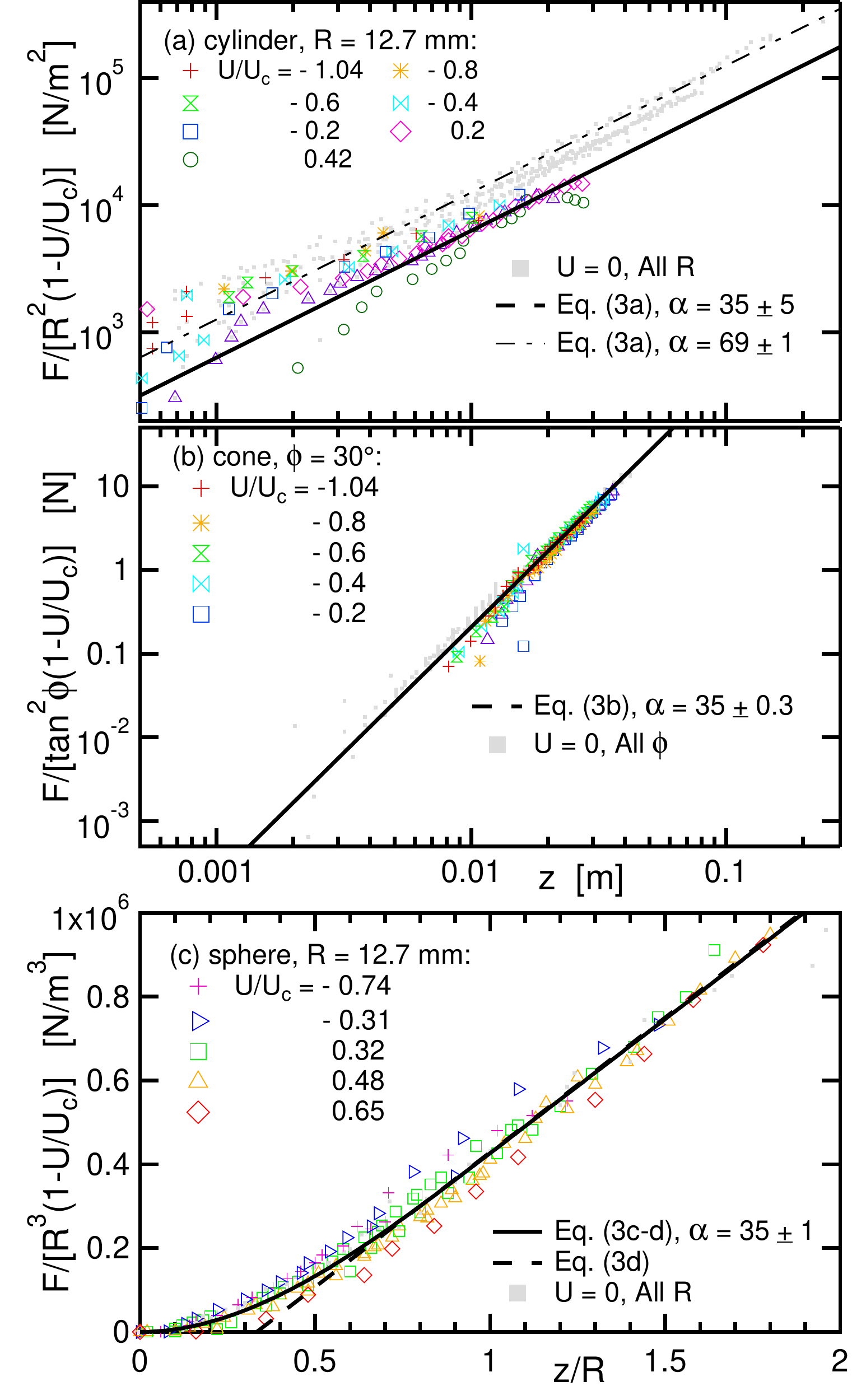}
\caption{(Color online) Friction force versus depth for (a)~cylindrical, (b)~conical, and (c)~spherical probes, normalized as in Fig.~\ref{fig:figure1} and also by $1-U/U_c$, where $U$ is airspeed and $U_c$ is airspeed at fluidization. The solid curves are simultaneous fits to Eq.~(\ref{eq:Shape}) with $\alpha=35\pm5$.  All data from Fig.~\ref{fig:figure1} are replotted as small grey points.}
\label{fig:figure2}
\end{figure}

The fits in Fig.~\ref{fig:figure2}(a-c) to Eq.~(\ref{eq:Shape}) are made using only one dimensionless parameter: $\alpha=35\pm5$. This value is close to $\alpha=26\pm 3$, as measured in \cite{katsuragi_unified_2007} for vertically impacting spheres at $U=0$, hinting that $\alpha$ may have a universal value close to 30 for vertical impacts onto cohesionless granular materials.  Between the quality of data collapse, the agreement between data and Eq.~(\ref{eq:Shape}), and the universality of the constant $\alpha$ for vertical impacts, we've demonstrated that the entirety of the effect of a sub-fluidizing flow of air through the sample is captured by the reduction factor $(1-U/U_{c})$. Thus the magnitude of $F(z)$ must be determined by gravity-loaded contacts in the bed, and not by contacts loaded via projectile motion.

As an aside, cylindrical projectiles in the absence of gas flow prove to be somewhat of a special case. The fit of Eq.~(\ref{eq:Cyl}) to the cylinder data in Fig.~\ref{fig:figure1}(a) gave $\alpha=70$, and is replotted as a dash-dotted line in Fig.~\ref{fig:figure2}.  So while the drag on cylinders exhibits radius- and depth-dependence consistent with Eq.~\ref{eq:Cyl}, it is double the magnitude of any other geometry in our quasistatic lowering experiments.  This is discussed further in \cite{supp}.

One might expect that the motion-loading to be more relevant for a projectile at $v\ne0$. In order to demonstrate that the behavior for quasi-static lowering is also valid more generally, we conduct several conventional impact experiments utilizing cylindrical and spherical projectiles. Each projectile is equipped with a small square rod, capped with a ferrous metal tip, to suspend the projectile from an electromagnet. We measure the height of the projectile above the granular surface, $h$, turn off the electromagnet, allow the projectile to free-fall onto the granular medium, and measure $d$ via height gauge. We repeat this procedure for a range of sub-fluidizing up- and down-flows of air to determine $d$ as a function of the normalized gas speed, $U/U_c$. 

The results for $d$ vs $U/U_c$, plotted in Fig.~\ref{fig:figure3}, conform to intuition. As $U\rightarrow U_c$, $d$ increases without apparent bound; and as $U$ becomes negative, $d$ decreases monotonically.  As a crucial check, penetration depths of the 12.7~mm radius sphere for three different granular packings are shown in (a) by large blue triangles: pointing up for usual conditions; pointing right for half the usual filling height, and hence half the imposed air pressure; pointing left for smaller grains, and hence a smaller $U_c$.  Despite the different conditions, the three data sets are indistinguishable.  This demonstrates the absence of both grain-size effects as well as interstitial air effects associated with packing fraction changes \cite{Goldman_PRE2010, Royer_Air}.  Thus, the up- or down-flow of air serves only to modify the gravity loading of contacts.

\begin{figure}
\includegraphics[width=2.50in]{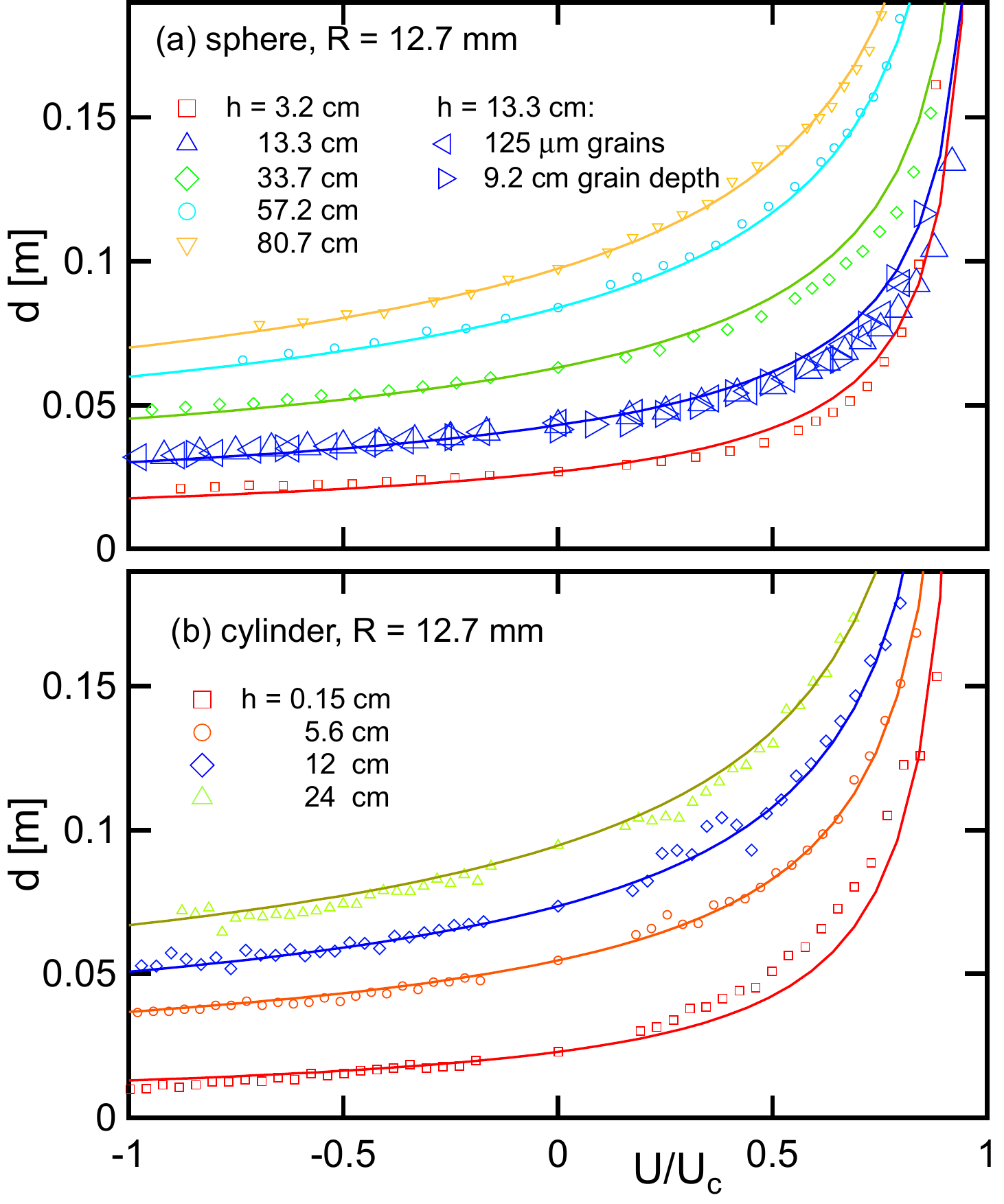}
\caption{(Color online) Penetration depth versus normalized airspeed for (a) a sphere of mass 0.07~kg, and (b) a cylinder of mass 0.2~kg, from various drop heights $h$.  The grains are 300~$\mu$m and the packings are 19.4~cm deep, except for two special cases in (a) where the depth and the grain size are decreased.  The curves represent solution of Eqs.~(\ref{eq:dFqs},\ref{eq:Cyl}) with $\alpha=35$, taking $b$ to match the data at $U=0$.}
\label{fig:figure3}
\end{figure}

Lastly we compare the measured final penetration depths with the exact solution of Eq.~(\ref{eq:TsimForce}), calculated in \cite{supp}.  For this, the initial impact speed is taken as $v=\sqrt{2gh}$, $F(z)$ is taken by Eq.~(\ref{eq:Cyl}) with $\alpha=35$, and $b$ is adjusted so that the prediction exactly matches the data at $U=0$.  The resulting predictions, shown as solid curves of Fig.~\ref{fig:figure3}, closely match the data for the entire range of airspeeds.  This demonstrates that the same rate-independent friction force acts during impact as during quasi-static lowering.


To conclude, the decoupling of frictional and inertial drag forces in Eq.~(\ref{eq:TsimForce}), seen earlier for horizontal motion \cite{brzinski_iii_characterization_2009}, also holds for downward motion.  The rate-independent upward friction term, $F(z)$, arises from local forces ${\rm d}{\bf F}$ according to Eq.~(\ref{eq:dFqs}) that point \emph{normal} to the intruder surface elements and that grow in proportion to a friction coefficient, gravitational hydrostatic pressure, and area.  While we varied the hydrostatic pressure via depth and airflow, and we varied area elements via intruder size and shape, we did not vary the friction coefficient.  However its presence is intuitive, and is supported by other work \cite{HiroakiPRE}.  It is surprising that friction does not act at motion-loaded contacts or have a significant component tangential to the intruder.  The latter can be judged somewhat just by plunging your finger into a container of grains.  This suggests that intruder roughness and tangential grain flow are not crucial, which is consistent with the observation that slick and tacky intruders have the same penetration depths \cite{uehara_low-speed_2003}.  This could be tested directly by roughening the intruder and imaging grain dynamics either for disks \cite{karen_2004, Clark_2012, Clark_2013} or index-matched grains~\cite{KN_impact}.

Another surprise is that the numerical coefficient in Eq.~(\ref{eq:dFqs}) is so large.  It is nice that $\alpha=35\pm5$ holds for nearly all data, but a number of order 1 would have been expected for ordinary Coulomb friction acting between grains and intruder.  Evidently, the relevant gravity-loaded contacts comprise a greater area and hence must be spread throughout a volume of grains near the intruder.  We propose a physical picture, based on force chains that extend from the intruder into the medium \cite{karen_2004, pica_ciamarra_dynamics_2004, Clark_2012, Clark_2013}.  These can be loaded statically and dynamically, and preferentially radiate away momentum \cite{Clark_2012, Clark_2013}.  They also tend to be oriented normal to the projectile, and to exist in a background sea of gravity-loaded grains.  Rigid-body motion of entire force chains could thus mobilize a large volume of gravity-loaded frictional grain-grain contacts, and give rise to a normal ${\rm d}{\bf F}$ and $\alpha\gg1$.  In effect, the chains are loaded between opposing forces from the intruder and from the surrounding gravity-loaded grains.  Perhaps this may also be thought of as a form of dynamical heterogeneity \cite{LucaDHbook}, whereby flow is accomplished by intermittently excited subpopulations of mobile and immobile grains.  Such an analogy correctly suggests that at very high intruder speeds, where grains are fluidized far from jamming, the stopping force is dominated by inertial effects, and vice-versa for very low intruder speeds.

\begin{acknowledgments}
This work was supported by the NSF through grant numbers DMR-0704147 and DMR-1305199.\end{acknowledgments}

\bibliography{CraterRefs}

\end{document}